\documentclass[a4paper,reqno]{amsart}
\usepackage[centertags]{amsmath}
\usepackage{amssymb}
\usepackage{amsthm}

\theoremstyle{plain}
\newtheorem{theorem}{Theorem}
\newtheorem{lemma}{Lemma}
\newtheorem{proposition}{Proposition}
\newtheorem{corollary}{Corollary}

\theoremstyle{definition}
\newtheorem{definition}{Definition}
\newcommand{\Dprime}[1]{\ensuremath{D'(#1)}}
\newcommand{\Rn}{\mathbf{R}^n}

\newcommand{\Ga}[2]{{\Gamma^#1}_{#2}}

\begin{document}
\title{Adiabatic Vacuumstates of the Dirac-field on a Curved Spacetime}
\author{Mark Wellmann}
\address{II.Institut f. Theoretische Physik \\
         Universit\"at Hamburg \\
         Luruper Chaussee 149 \\
         D-22761 Hamburg}
\email{wellmann@x4u2.desy.de}

\begin{abstract}
In this article we review the quantization of the Dirac-field on a curved
spacetime. For that purpose we describe the construction of the local observable
algebras in the algebraic approach to quantum field theory. Among the possible
states we single out the so called Hadamard-states, which are the ones relevant
for physics. Finally, as an example, we give a definition for an adiabatic
vacuum state of the Dirac-field on a Robertson-Walker spacetime. We believe that
these states are physical in the sense that they have the singularity structure
of Hadamard-form, although we cannot give a formal proof of this conjecture.
\end{abstract}

\maketitle

\section{Introduction}
In this paper we are concerned with quantum field theory on a curved spacetime
(QFT on CST). Among the vast literature in this area on the Klein-Gordon-field
only little work is done for the Dirac-field. The reason for this are not
conceptual problems but the greater technical complexity involved in the
description of multicomponent fields.

After Hawkings discovery that a black hole not only absorbs the energy incident
on it, but also loses energy through the emission of pairs of elementary
particles there were a great interest in the treatment of quantized fields in a
classical background. This semiclassical description of quantum fields
propagating in a non flat background can serve as an approximation to the yet
non existing theory of quantum gravity. The range of validity goes down to the
Planck-length, where quantum gravity effects are expected to become important.

One of the main questions in this branch was concerned with the characterization
of physical states on the various spacetime models. It is now accepted that all
physical quantum states must have a singularity structure of Hadamard-form. For
these states there is a well defined prescription to regularize the
energy-momentum tensor and the semiclassical Einstein-equations
\begin{equation*}
  G_{ab}=8\pi\langle T_{ab}\rangle
\end{equation*}
which govern the backreaction of the quantized fields on the spacetime geometry
make sense. Recently there was made progress in the formulation of the
spectrum condition on a curved spacetime. There the Hadamard-states play an important
role too, see~\cite{bfk:msc}.
\section{Dirac-field}
To set our notation we give in this section a short description of spinors in
arbitrary curved spacetimes. Furthermore we recall the existence- and
uniqueness-results of fundamental solutions of the Dirac-equation. At the end we
mention Dimock's result that the Cauchy-Problem is well posed~\cite{di:dqf}.

By curved spacetime we mean a 4-dimensional $\mathcal{C}^{\infty}$-manifold
which is endowed with a metric, a covariant tensor-field of type (0,2), with
Lorentzian signature $(+,-,-,-)$. The metric describes locally the causal
structure of our spacetime. We first introduce the complex Clifford-algebra of
Dirac-matrices on Minkowski-spacetime and then we use the tetrad-formalism to
lift them on our spacetime manifold. We denote the Minkowski-spacetime by
$(\mathbf{R}^4,\eta_{ab}=\text{diag}(1,-1,-1,-1))$. Let
$\mathcal{C}(\mathbf{R}^4,\eta)$ be the real, associative algebra with unit
$\mathbf{1}$ which is generated by the elements $\{c(v), v\in \mathbf{R}^4\}$
and the relation
\begin{equation}
 \{c(v),c(w)\}=2\eta(v,w)\mathbf{1}                                                         
\end{equation}
This is the Clifford-algebra associated to Minkowski-spacetime
$(\mathbf{R}^4,\eta_{ab})$. In the description of the Dirac-field on
Minkowski-spacetime the so called Dirac-matrices play a prominent role. We
obtain these matrices when we look at a faithful representation $\rho$ of the
complexified Clifford-algebra $\mathcal{C}(\mathbf{R}^4,\eta)_{\mathbf{C}}$ on
$M_{\mathbf{C}}(4)$. The resulting Dirac-matrices are denoted by
$\gamma_a:=\rho(c(e_a)),\,a=0,\ldots 3$. The vectors $\{e_a\}$ built a standard
basis of $\mathbf{R}^4$. For two different representations $\rho_1$ and $\rho_2$
of $\mathcal{C}(\mathbf{R}^4,\eta)_{\mathcal{C}}$ we can always find a
nonsingular matrix $T$ which transforms the two corresponding sets of
Dirac-matrices $\gamma^1_a$ and $\gamma^2_a$ into each other, i.e.
\begin{equation}
  T\gamma^1_aT^{-1}=\gamma^2_a,\ a=0,\ldots, 3                                              
\end{equation}
In the following we assume a form of the Dirac-matrices with the additional
property
\begin{equation}
  \gamma^{\ast}_0=\gamma_0,\ \gamma^{\ast}_j=-\gamma_j                                      
\end{equation}
The $\ast$ means hermitian conjugation. Dirac-matrices with this property are
said to be in standard representation. Next we consider the Lie-group
$\text{Spin}(1,3)$ of matrices $S$ with the properties
\begin{align}
  \text{det}S     & = 1 \\                                                                  
  S\gamma_aS^{-1} & = \gamma_b{\Lambda^b}_a                                                 
\end{align}
Together with the anticommutation relations of the Dirac-matrices one sees that
the matrix $\Lambda$ must be an element of the full Lorentz-group $\mathcal{L}$.
The mapping $S\rightarrow \Lambda(S)$ defined by (5) is a 2-1 homomorphism of
$\text{Spin}(1,3)$ onto $\mathcal{L}$. In the following we restrict ourselves to
$\text{Spin}_0(1,3)$ the connected component of the identity of
$\text{Spin}(1,3)$. The image under the above mapping is the group of proper
orthochronous Lorentz transformations $\mathcal{L}^{\uparrow}_+$.

Now let $(M,g)$ the spacetime under consideration. We suppose $(M,g)$ to be
time- and space-orientable, which means that we can make a continuous distinction
between future and past as well as left and right throughout $M$, and choose
such orientations. Let $TM$ denote the tangent bundle over $M$. To a given
coordinate basis $\{e_{\mu},\,\mu=0,\ldots 3\}$ at the point $p\in M$ we
associate an orthonormal basis $\{e_a,\,a=0,\ldots 3\}$ which is a linear
combination of the coordinate basis vectors $e_{\mu}$ according to
\begin{align}
  e_a                                 & = {e^{\mu}}_a e_{\mu}\\                             
  g(e_a,e_b)                          & = \eta_{ab}\\                                       
  g_{\mu\nu}(p){e^{\mu}}_0{e^{\nu}}_0 & \geq 0,\,{e^0}_0\geq 0,\ p\in M                     
\end{align}
The matrix $({e^{\mu}}_a)$ shall be an element of $GL(4,\mathbf{R})$ and the
last condition is necessary to preserve the orientations. The collection of all
this orthonormal bases constitute the so called orthonormal frame bundle
$F(M,g)$ associated to $TM$. Under a change of frame the tetrad ${e^{\mu}}_a$
transforms as
\begin{equation}
  {{e^{\prime}}^{\mu}}_b={e^{\mu}}_a{\Lambda^a}_b,\ \Lambda\in \mathcal{L}^{\uparrow}_+     
\end{equation}
If we perform a local change of coordinates around $p\in M$ we obtain the
following transformation law for the tetrad
\begin{equation}
  {e^{\mu}}_a=\left( \frac{\partial x^{\mu}}{\partial y^{\nu}}
  \right)_p{{e^{\prime}}^{\nu}}_a                                                          
\end{equation}
In this way $F(M,g)$ admits the structure of a
$\mathcal{L}^{\uparrow}_+$-principal fiberbundle. If we consider
$\text{Spin}_0(1,3)$ instead of $\mathcal{L}^{\uparrow}_+$ we obtain a
$\text{Spin}_0(1,3)$-principal fiberbundle which is called a spin structure
$S(M,g)$ for $(M,g)$, if we have in addition a bundlehomomorphism
$\varphi:S(M,g)\rightarrow F(M,g)$ which satisfies
\begin{equation}
  \varphi\circ R_S=R_{\Lambda(S)}\circ \varphi,\ S\in\text{Spin}_0(1,3)                    
\end{equation}
The symbol $R$ denotes the right action with $\Lambda$ resp. $S(\Lambda)$ on
orthonormal basis vectors in $F(M,g)$ resp. $S(M,g)$. If a given spacetime
possesses spin structures is determined by the topological properties of the
underlying manifold. Here we consider only globally hyperbolic spacetimes for
which the manifold is so well behaved that the existence of spin structures is
always guaranteed, but they are not uniquely determined by the topology.

Now we are ready to define Dirac-spinors. On our spacetime $(M,g)$ with spin
structure $(S(M,g),\varphi)$ we take a standard representation $\rho$ of the
Clifford-algebra $\mathcal{C}(\mathbf{R}^4,\eta)_{\mathcal{C}}$ together with
the group $\text{Spin}_0(1,3)$. The bundle of Dirac-spinors is now the
associated vectorbundle $D_{\rho}(M,g)$ to the $\text{Spin}_0(1,3)$-principal
fiberbundle $S(M,g))$, the representation $\rho$ and the vectorspace
$\mathbf{C}^4$:
\begin{equation}
  D_{\rho}(M,g):=(S(M,g)\times_{\rho}\mathbf{C}^4)/\text{Spin}_0(1,3)                      
\end{equation}
In the following we abbreviate $D_{\rho}(M,g)$ with $DM$. A spinorfield is now a
smooth section in $DM$. The collection of all smooth spinorfields is denoted by
$\mathcal{C}^{\infty}(DM)$. Those which have in addition compact support are
denoted by $\mathcal{C}^{\infty}_0(DM)$. With $D^{\ast}M$ we denote the dual
spinorbundle. A $v\in D^{\ast}M$ is for $p\in M$ an element of the dual space to
the fiber over $p$. The smooth sections in $D^{\ast}M$ are called
co-spinorfields. The set of all smooth co-spinorfields are denoted by
$\mathcal{C}^{\infty}(D^{\ast}M)$. Let $\Sigma$ be a submanifold of $M$. By
$DM_{\Sigma}$ we denote the vectorbundle $\pi^{-1}_{DM}(\Sigma)$ over $\Sigma$
with projection $\pi_{DM|\Sigma}:=\pi_{DM}|\pi^{-1}_{DM}(\Sigma)$, where
$\pi_{DM}$ is the projection in $DM$. For a spinor $u\in
\mathcal{C}^{\infty}(DM)$ and a co-spinor $v\in
\mathcal{C}^{\infty}(D^{\ast}M)$ we have a natural dual pairing
\begin{equation}
  v(u)|_p=v_Au^A,\ p\in M                                                                  
\end{equation}
with respect to bases $\{E_A,\,A=0,\ldots 3\}$ resp. $\{E^A,\,A=0,\ldots 3\}$ in
$\mathcal{C}^{\infty}(DM)$ resp. $\mathcal{C}^{\infty}(D^{\ast}M)$. In the above
expression we use the Einstein summation convention. We can now generate
arbitrary spinor-tensor fields, i.e. smooth sections in the fiberwise tensor
products of $TM, T^{\ast}M, DM, D^{\ast}M$. An element $f$ in
$(\bigotimes_pTM)\otimes(\bigotimes_qT^{\ast}M)\otimes(\bigotimes_rDM)\otimes
(\bigotimes_sD^{\ast}M)$ is given by specifying the family of
$\mathbf{C}$-valued functions ${{{f^{a_1\ldots a_p}}_{b_1\ldots b_q}}^{A_1\ldots
A_r}}_{B_1\ldots B_s}$ on $M$, such that
\begin{equation}
  f=({{{f^{a_1\ldots a_p}}_{b_1\ldots b_q}}^{A_1\ldots A_r}}_{B_1\ldots
  B_s})e_{a_1}\otimes
  \ldots \otimes e_{a_p}\otimes \ldots \otimes E^{B_1}\otimes \ldots \otimes E^{B_s}       
\end{equation}
Of special interest is the spinor-tensor $\gamma \in T^{\ast}M\otimes DM\otimes
D^{\ast}M$ which has the components
\begin{equation}
  {{\gamma_a}^A}_B={{(\gamma_a)}^A}_B                                                      
\end{equation}
with the Dirac-matrices $\gamma_a,\,a=0,\ldots 3$ in standard representation.
For a vector field $k \in \mathcal{C}^{\infty}(TM)$ we denote by ${\not k}$ the
contraction of $k$ with $\gamma$, i.e. ${{\not k}^A}_B:=k^a{{\gamma_a}^A}_B$.
Next we introduce the notion of an adjoint spinor $u^+$ for a spinor $u$. This
is the co-spinor with the following components:
$(u^+)_B:=\overline{u^A}\gamma_{0AB}$, where the bar means complex conjugation.
We can now use the spinor-tensor $\gamma$ to lift the
Levi-Civit$\acute{\text{a}}$-derivative of the metric $g$ to mixed spinor-tensor
fields in the following way. The usual covariant derivative $\nabla
:\mathcal{C}^{\infty}(TM)\rightarrow \mathcal{C}^{\infty}(T^{\ast}M\otimes TM)$
is defined by ${(\nabla k)^b}_a:=\partial_ak^b+{\Gamma^b}_{ac}k^c,\,k\in
\mathcal{C}^{\infty}(TM)$. We extend this definition to spinor fields as follows:
$\nabla:\mathcal{C}^{\infty}(DM)\rightarrow\mathcal{C}^{\infty}(T^{\ast}M\otimes
DM)$ is defined through the specification of the components of $\nabla
f,\,f\in\mathcal{C}^{\infty}(DM)$ with respect to bases $(E^A)$ and $(e_a)$ as
${(\nabla f)_a}^B:=\partial_af^B+{{\sigma_a}^B}_Af^A$, where the so called spin
connection coefficients are given by
\begin{equation}
  {{\sigma_a}^B}_A:=-\frac{1}{4}{\Gamma_{ca}}^b{{\gamma_b}^B}_D{\gamma^{cD}}_A             
\end{equation}
This defines indeed a covariant derivative on $M$. In demanding the
Leibnitz-rule and commutativity with contractions, we extend this definition to
all spinor-tensor fields. As an easy consequence of the above definitions we
obtain the following lemma
\begin{lemma}
  The spinor-tensor $\gamma$ is covariant constant, i.e. $\nabla\gamma=0$.
\end{lemma}
\begin{proof}
One can see this immediately by using the anticommutation relations of the Dirac-matrices
\renewcommand{\qed}{\hfill$\blacksquare$}
\end{proof}

We can now write down the Dirac-equations for spinors and cospinors, which are
\begin{align}
  (-i{\not \nabla}+m)u & = 0,\,u\in \mathcal{C}^{\infty}(DM)\\                             
  (i{\not \nabla}+m)v  & = 0,\,v\in \mathcal{C}^{\infty}(D^{\ast}M)                        
\end{align}
In these equations $m$ is a fixed real number. The Dirac-operator obeys
Lichnerowicz' identity:
\begin{equation}
  (-i{\not \nabla}+m)(i{\not \nabla}+m)u=(\square -\frac{1}{4}R+m^2)u,\,u\in
  \mathcal{C}^{\infty}(DM)                                                                 
\end{equation}
where $\square =\eta^{ab}\nabla_a\nabla_b$ is the spinorial wave operator. The
operator on the right side of (19) is known as the spinorial
Klein-Gordon-operator. Next we consider the classical solutions to the
Dirac-equation on a globally hyperbolic spacetime. We remind ourselves that a
spacetime with given time- and space orientations is said to be globally
hyperbolic, if it admits a Cauchy-surface. A Cauchy-surface is a smooth
spacelike hypersurface $\Sigma\subset M$ which is intersected by every endless
causal curve exactly once. A globally hyperbolic spacetime has the structure
$\mathbf{R}\times\Sigma$, i.e we can foliate $M$ into smooth hypersurfaces. We
now introduce distributions on the spaces $\mathcal{C}^{\infty}_0(DM)$ and
$\mathcal{C}^{\infty}(D^{\ast}M)$. The natural pairing $v(u)$ between
$v\in\mathcal{C}^{\infty}(D^{\ast}M)$ and $u\in \mathcal{C}^{\infty}_0(DM)$ can
be integrated over $M$ in the obvious way:
\begin{equation}
  \langle v,u\rangle_M:=\int_Mv(u)(p)d\mu(p),\,u\in
  \mathcal{C}^{\infty}_0(DM),\, v\in \mathcal{C}^{\infty}(D^{\ast}M)                       
\end{equation}
Here $d\mu(p)$ is the volume element associated with the metric $g$. In addition
to that we demand continuity of $v$, seen as a distribution over
$\mathcal{C}^{\infty}_0(DM)$, with respect to certain Sobolev-norms, see
\cite{ver:sa} for details. In this way we have an embedding of
$\mathcal{C}^{\infty}(D^{\ast}M)$ in $\mathcal{C}^{\infty}_0(DM)^{\prime}$, the
space of cospinor-valued distributions over $M$. Analogously
$\mathcal{C}^{\infty}_0(DM)$ is embedded in the space of spinor-valued
distributions over $M$ with compact support,
$\mathcal{C}^{\infty}(D^{\ast}M)^{\prime}$. The next Theorem gives us the
existence of unique fundamental solutions to the Dirac operators
\begin{theorem}[Dimock, 1982]
  The operator $(-i{\not \nabla}+m)$ on $\mathcal{C}^{\infty}(DM)$ has unique retarded
  resp. advanced fundamental solutions $S^{\pm}:\mathcal{C}^{\infty}_0(DM)\rightarrow
  \mathcal{C}^{\infty}(DM)$ with
  \begin{equation}
    (-i{\not \nabla}+m)S^{\pm}=S^{\pm}(-i{\not \nabla}+m)=
    \text{id on}\ \mathcal{C}^{\infty}_0(DM)                                               
  \end{equation}
  and supp$(S^{\pm}f)\subset\text{J}^{\pm}
  (\text{supp}(f))$.\footnote{Here for $N\subset M\,\text{J}^{\pm}(N)$ is the set of
  points in $M$, which are connected to $N$ through a future directed (+) resp. past
  directed (-) causal curve.}
  The operator $(i{\not \nabla}+m)$ on $\mathcal{C}^{\infty}(D^{\ast}M)$ has
  unique retarded resp. advanced fundamental solutions $S_{\pm}:
  \mathcal{C}^{\infty}_0(D^{\ast}M)\rightarrow \mathcal{C}^{\infty}(D^{\ast}M)$ with
  \begin{equation}
    (i{\not \nabla}+m)S_{\pm}=S_{\pm}(i{\not \nabla}+m)=\text{id on}\
    \mathcal{C}^{\infty}_0(D^{\ast}M)                                                      
  \end{equation}
   and supp$(S_{\pm}f)\subset \text{J}^{\pm}(\text{supp}(f))$.
\end{theorem}
The difference $S$ between retarded $(S^+)$ and advanced $(S^-)$ fundamental
solution is called Dirac-Propagator for spinors: $S=S^+-S^-$.
$S_{\sharp}=S_+-S_-$ is called Dirac-Propagator for cospinors.

\noindent The Cauchy problem for the Dirac-equation is now well posed:
\begin{theorem}[Dimock, 1982]
  Let $u_0\in\mathcal{C}^{\infty}_0(DM_{\Sigma})$. Then there is a unique
  $u\in \mathcal{C}^{\infty}(DM)$ with
  \begin{equation}
    (-i{\not \nabla}+m)u=0\quad\text{and}\quad u|_{\Sigma}=u_0                             
  \end{equation}
  In addition to that we have supp$(u)\subset\text{J}^+(\text{supp}(u_0))
  \cup\text{J}^-(\text{supp}(u_0))$.
\end{theorem}
These two theorems lie at the heart of the quantization of the Dirac-field which
is described in the next section.

\section{Quantization}
In this section the quantization of the Dirac-field on an arbitrary curved
spacetime is described. Since on a general curved spacetime we do not have time
translation symmetry, we have no natural criterion to single out a Hilbertspace
of solutions, oscillating with positive frequency, which play the role of a
one-particle space in Minkowski-spacetime. Therefore we adopt the algebraic
approach here, in which only the algebraic structure of the theory is specified,
i.e. we construct a $\mathcal{C}^{\ast}$-algebra and in a subsequent step
certain subalgebras which are associated to open regions in spacetime~\cite{wa:qft}. These
local algebras constitute the so called observable algebras in the sense of Haag
and Kastler~\cite{hk:aaq}. A state is then specified as a positive normalized
linear functional on the observables. The connection to the usual operator
approach is given by the GNS-construction.

Let $\mathsf{S}$ be the space of smooth complex-valued solutions to the
Dirac-equation, which have compact support initial data:
\begin{equation}
  \mathsf{S}=\{u\in\mathcal{C}^{\infty}(DM)|(-i{\not\nabla}+m)u=0\ \text{and}\             
  u|_{\Sigma}=u_0\in\mathcal{C}^{\infty}_0(DM_{\Sigma})\}
\end{equation}
On $\mathsf{S}$ we introduce the following scalar product
$(.\,,.):\mathsf{S}\times\mathsf{S}\rightarrow\mathbf{C}$, given by
\begin{equation}
  (u_1,u_2):=\int_{\Sigma}\tilde{u_1}^+({\not n}\tilde{u_2})(p)
  d\mu_{\Sigma}(p)                                                                         
\end{equation}
In this expression $\tilde{u}$ is the restriction of the spinorfield $u$ to
$\Sigma$. $n$ is the unit normal vectorfield on $\Sigma$ and $\mu$ is the volume
element induced by the three-dimensional Riemannian metric
$h_{ab}=g_{ab}+n_an_b$ on $\Sigma$. This scalar product is independent of the
choice of Cauchy-surface $\Sigma$. We complete the space $\mathsf{S}$ in the
norm induced by this scalar product and obtain in this way a complex
Hilbertspace $\mathsf{H}$. Now we can associate to $\mathsf{H}$ an abstract
$\mathcal{C}^{\ast}$-algebra $\mathcal{F}[\mathsf{H}]$ by the antilinear mapping
$B:\mathsf{H}\rightarrow\mathcal{F}[\mathsf{H}]$ such that
$\{B(h):h\in\mathsf{H}\}$ together with the unit $\mathbf{1}$ generate
$\mathcal{F}[\mathsf{H}]$ algebraically and fulfill the canonical
anticommutation relations:
\begin{align}
  \{\,B(h)\,,\,B(g)\,\}        & = 0 \\                                                    
  \{\,B(h)\,,\,B(g)^{\ast}\,\} & = \ (h,g)\cdot\mathbf{1},\ \text{for all}\
                                   h,g\in\mathsf{H}                                        
\end{align}
The algebra $\mathcal{F}[\mathsf{H}]$ has a unique $\mathcal{C}^{\ast}$-norm
$||\,.\,||_{\mathcal{C}^{\ast}}$. Moreover it follows by the use of the
CAR-relations that the mapping $B$ is an isometry, i.e. we have
\begin{equation}
  ||B(f)||_{\mathcal{C}^{\ast}}=||f||,\ f\in\mathsf{H}                                     
\end{equation}
By assigning to every open relative compact subset $\mathcal{O}\subset M$ the
$\mathcal{C}^{\ast}$-subalgebra $\mathcal{F}(\mathcal{O})$ which is generated by
elements $\{B(f):f\in\mathcal{C}^{\infty}_0(DM,\mathcal{O})\}$, where we denote
by $\mathcal{C}^{\infty}_0(DM,\mathcal{O})$ spinors whose support lies in
$\mathcal{O}$, we obtain a net of field algebras
$\mathcal{O}\rightarrow\mathcal{F}(\mathcal{O})$. The even part hereof, i.e.
elements of the form
$\{B(h)^{\ast}B(f):h,f\in\mathcal{C}^{\infty}_0(DM,\mathcal{O})\}$ built the net
of observable algebras $\mathcal{O}\rightarrow \mathcal{A}(\mathcal{O})$ in the
sense of Haag and Kastler~\cite{hk:aaq}.

Now we come to the consideration of states on our CAR-algebra
$\mathcal{F}[\mathsf{H}]$. A state is a positive, normalized linear functional
$\omega$ on $\mathcal{F}[\mathsf{H}]$, by which we mean a linear map
$\omega:\mathcal{F}[\mathsf{H}]\rightarrow\mathbf{C}$ with the properties
\begin{align}
  \omega(B(f)^{\ast}B(f)) & \ \geq 0\quad \text{(positivity)}\\                            
  \omega(\mathbf{1})      & = 1\quad \text{(normalization)}                                
\end{align}
As usual we focus attention on quasifree states, which have the form
\begin{equation}
  \omega(B(f_1)\ldots B(f_{2n}))=(-1)^{n(n-1)}\sum\text{sgn}(\sigma)\prod_{j=1}^n
  \omega(B(f_{\sigma(j)})B(f_{\sigma(j+n)})),\,n\in \mathbf{N}                             
\end{equation}
where the sum runs through all permutations $\sigma$ of $\{1,\ldots,2n\}$ with
$\sigma(1)<\sigma(2)<\ldots<\sigma(n)$ and
$\sigma(j)<\sigma(j+n),\,j=1,\ldots,n$. sgn$(\sigma)$ is $+1(-1)$, if $\sigma$
is an even (odd) permutation of $\{1,\ldots,2n\}$. In addition to that $\omega$
vanishes on all odd monomials $B(f_1)\ldots B(f_{2n+1}), n\in\mathbf{N}_0$.

Let $P$ be an orthogonal projection in $\mathsf{H}$. A quasifree state
$\omega_{P}$ on $\mathcal{F}[\mathsf{H}]$ can then be specified in the following
way. We choose an orthonormal basis $\{f_i\}_{i\in\mathbf{N}}$ for $\mathsf{H}$
consisting of eigenspinors for $P$. Then every element $h\in\mathsf{H}$ has an
expansion, $h=\sum_{i\in\mathbf{N}}(f_i,h)f_i$. We define an operator $\Gamma$
on $\mathsf{H}$ by $\Gamma h=\sum_{i\in\mathbf{N}}
\overline{(f_i,h)}f_i$. $\Gamma$ is a well defined involution and it is
antiunitary, i.e.
\begin{equation}
  (\Gamma g,\Gamma h)=(h,g),\quad \forall g,h\in\mathsf{H}                                 
\end{equation}
Moreover $\Gamma$ commutes with $P$. Then the spaces $P\mathsf{H}$ and
$(I-P)\mathsf{H}$ together span the entire Hilbertspace $\mathsf{H}$ and are
orthogonal w.r.t. $(.\,,.)$. Therefore we can write every element
$h\in\mathsf{H}$ in a unique way as
\begin{equation}
  h=h^++h^-,\quad h^+\in P\mathsf{H},\,h^-\in (I-P)\mathsf{H}                              
\end{equation}

Now we take the antisymmetric Fockspace $\mathcal{F}_{a}(\mathsf{H})$ over
$\mathsf{H}$ with annihilation operator $a(f):\wedge^{n+1}\mathsf{H}\rightarrow
\wedge^n\mathsf{H}$ defined by
\begin{equation}
  a(f)\Omega=0                                                                             
\end{equation}
and creation operator $a(f)^{\ast}:\wedge^n\mathsf{H}\rightarrow
\wedge^{n+1}\mathsf{H}$ given by
\begin{align}
  a(f)^{\ast}\Omega                      & = f \\                                          
  a(f)^{\ast}(f_1\wedge\ldots\wedge f_n) & = f\wedge f_1\wedge\ldots\wedge f_n,\           
                                             \forall f,f_1,\ldots, f_n\in\mathsf{H}
\end{align}
where $\Omega$ is the vacuum vector in $\mathcal{F}_{a}(\mathsf{H})$. With
$\wedge^n\mathsf{H}$ we denote the antisymmetric tensor product
$\mathsf{P}_{\wedge}(\otimes^n\mathsf{H})$ defined by
\begin{equation*}
  \mathsf{P}_{\wedge}(f_1\otimes\ldots\otimes f_n)=(n!)^{-1}\sum_{\sigma\in S_n}\chi_{\wedge}
  (\sigma)\cdot f_{\sigma(1)}\otimes\ldots\otimes f_{\sigma(n)}
\end{equation*}
where the sum runs through all permutations $\sigma$ in the permutation group
$S_n$ and $\chi_{\wedge}$ is the sign of the permutation $\sigma$. Furthermore
we have used the notation
\begin{equation*}
  f_1\wedge\ldots\wedge f_n=(n!)^{\frac{1}{2}}\mathsf{P}_{\wedge}(f_1\otimes\ldots\otimes
  f_n)
\end{equation*}
The operators $a(f)$ and $a(f)^{\ast}$ are bounded operators on
$\mathcal{F}_{a}(\mathsf{H})$ and they have anticommutation relations given by
\begin{equation}
  \{\,a(f)\,,\,a(g)^{\ast}\,\}=(f,g)\cdot\mathbf{1}                                        
\end{equation}
This is a concrete $\mathcal{C}^{\ast}$-algebra, the so called
Fock-representation of the CAR. We take the Dirac-field operator to be
\begin{equation}
  \Psi_P(f)=a(f^+)+a(\Gamma f^-)^{\ast},\ f=f^++f^-\ \text{where}\ f^+\in P\mathsf{H},\,   
  f^-\in(I-P)\mathsf{H}
\end{equation}
Then we have the anticommutation relation
\begin{equation}
  \{\,\Psi(f)\,,\,\Psi(g)^{\ast}\,\}=(f,g)\cdot\mathbf{1}                                  
\end{equation}
This is known as the quasifree representation of the CAR-algebra
$\mathcal{F}[\mathsf{H}]$. The state $\omega_P$ is the quasifree state
determined by the two-point function
\begin{equation}
  \Lambda^{\omega_P}_2(f,g)=\langle\Omega,\Psi(f)\Psi(g)^{\ast}\Omega\rangle               
\end{equation}
where $\langle.\,,.\rangle$ is the inner product in
$\mathcal{F}_{a}(\mathsf{H})$.

\section{Hadamard states}
In this section we are looking at those states for which the two-point function
has the singularity structure of Hadamard form~\cite{dwb:rdgf}. For these states it has been
proven in the case of the Klein-Gordon field on a curved spacetime that the
resulting operator description, which was sketched above, gives a physical
theory in the sense that the principle of local definiteness is
satiesfied~\cite{hns:qftgb,ver:ldpq}. We define now Hadamard states for the
Dirac-field which are believed to be equally important here as they are for the
scalar case. For the introduction of these states we need the concept of
bi-spinors and some technicalities concerning the causal structure of spacetime
which we relegate in an appendix.

Let $(M,g)$ be a globally hyperbolic spacetime which Cauchy-surface $\Sigma$.
The square root of the bi-spinor valued VanVleck-Morette determinant is defined
by
\begin{equation}
  {U^A}_{B'}=\Delta^{1/2}(q,q'){\mathcal{J}^A}_{B'}                                        
\end{equation}
where $\Delta^{1/2}(q,q')=g(q)^{1/2}\text{det}(\sigma(q,q')_{;ab})g(q')^{1/2}$
is the scalar VanVleck-Morette determinant and $\sigma(q,q')$ is one half of the
square of the geodesic distance $s(q,q')$. The bi-spinor of the parallel
transport is defined by
\begin{align}
  \sigma^{;\mu}{\mathcal{J}^A}_{B';\mu} & = 0 \\                                           
  {\mathcal{J}^A}_{B'}(q,q)             & = \ {\mathbf{1}^A}_{B'}                          
\end{align}
Let ${{V_m}^A}_{B'},\,m\in\mathbf{N}_0$ be a sequence of bi-spinors which are
determined by the Hadamard recurrence relations~\cite{kh:set}. Both ${U^A}_{B'}$
and ${{V_m}^A}_{B'}$ are well defined on $X$, the set of points which can be
joined by a unique geodesic (see the appendix for a precise definition of $X$).
Furthermore we define for all $n\in \mathbf{N}$
\begin{equation}
  {V^{(n)A}}_{B'}(q,q'):=\sum_{m=0}^n{{V_m}^A}_{B'}(q,q')(s(q,q'))^m,\,(q,q')\in X         
\end{equation}
Finally we take a global time function $T:M\rightarrow \mathbf{R}$ and set
\begin{equation}
  r_{\epsilon}^T(q,q'):=-2i\epsilon(T(q)-T(q'))-\epsilon^2,\,q,q'\in
  M,\,\epsilon>0                                                                           
\end{equation}
When we look at the Dirac-operator we observe that it is not a hyperbolic
operator in the sense of Leray~\cite{le:hde}. We use the fact that, due to
Lichnerowicz' identity, the product of the Dirac-operator times it's complex
conjugate is an hyperbolic operator. In the next definition we will introduce
the singularity structure of Hadamard form for the spinor Klein-Gordon operator
which underlies the definition of Hadamard state for the Dirac-field.
\begin{definition}
  Let $\Lambda$ be a sesquilinear form on $\mathcal{C}^{\infty}_0(DM)$. $\Lambda$ has
  Hadamard form (for the spinor Klein-Gordon operator), if the following data exist:
  \begin{enumerate}
    \item  a causal normal neighborhood $N$ of a Cauchy-surface $\Sigma$
    \item  a $N$-regularising function $\chi$
    \item  a smooth future directed global time function $T$ on $M$
    \item  a sequence $H^{(n)}\in\mathcal{C}^n(DN\boxtimes D^{\ast}N),\,n\in\mathbf{N}$, so
               that for all $n\in\mathbf{N}$ and all $f,h\in\mathcal{C}^{\infty}_0(DM,N)$
               \begin{equation*}
                 \Lambda(f,h)=\lim_{\epsilon\rightarrow 0}\int{{(\Lambda^{T,n}_{\epsilon})}
                 ^A}_{B'}(q,q')f^+_A(q)h^{B'}(q')d\mu(q)d\mu(q'),\,\text{where}
               \end{equation*}
               \begin{equation*}
                 {(\Lambda^{T,n}_{\epsilon})^A}_{B'}(q,q'):=\chi(q,q')
                 {{(G^{T,n}_{\epsilon})}^A}_{B'}(q,q') + {H^{(n)A}}_{B'}(q,q')\ \text{with}
               \end{equation*}
               \begin{equation*}
                 \begin{split}
                   {{(G^{T,n}_{\epsilon})}^A}_{B'}(q,q') := & \ \frac{1}{4\pi^2} \Bigl(
                                                              \frac{{U^A}_{B'}(q,q')}{s(q,q')+
                                                              r_{\epsilon}^T(q,q')}+
                                                              {V^{(n)A}}_{B'}(q,q')\ln(s(q,q')\\
                                                            & +r_{\epsilon}^T(q,q')) \Bigr)
                 \end{split}
               \end{equation*}
               for $(q,q')\in X\cap (N\times N)$; $\ln$ is the main branch of the logarithm
  \end{enumerate}
\end{definition}
We now view the two-point function $\omega(B(f_1)^{\ast}B(f_2))$ as a hermitian
form $f_1,f_2\mapsto (f_1,Qf_2)$ over $\mathsf{H}$. This form is a solution of
the Dirac equation in both arguments, i.e. we have
\begin{equation}
  ((-i{\not\nabla}+m)f_1,Qf_2)=(f_1,Q(-i{\not\nabla}+m)f_2)=0                              
\end{equation}
With the next definition we arrive at the Hadamard form of a state of the
Dirac-field
\begin{definition}
  A quasifree state $\omega$ on $\mathcal{F}[\mathsf{H}]$ is called a Hadamard state of the
  Dirac-field, if there is a solution $\Lambda$ of the spinorial Klein-Gordon equation of
  Hadamard form, so that the two-point function of $\omega$ has the following form
  \begin{equation}
    \omega(B(f_1)^{\ast}B(f_2))=\Lambda((i{\not\nabla}+m)f_1,f_2),\,
    f_1,f_2\in\mathsf{H}                                                                   
  \end{equation}
\end{definition}
$\Lambda$ is a solution of the Dirac equation in the first argument, since it
fulfills Lichnerowicz' identity and in the second argument because of
hermiticity. The singular part of $\Lambda$ is determined by geometric
quantities like metric and spinor connection only. Therefore a state is
characterized, if we fix the smooth part of $\Lambda$.

Since the work of Radzikowski~\cite{rad:hc} there is another possibility to
characterize the singularity structure of a Hadamard state. This method uses
H\"ormanders wavefront sets~\cite{hm:fio1} to describe the singular behavior of
the numerical distributions
\begin{equation}
  f_1,\ldots, f_n\mapsto\Lambda^{\omega}_n(f_1,\ldots, f_n):=\langle\Omega,
  \Psi(f_1)\ldots\Psi(f_n)\Omega\rangle                                                    
\end{equation}
We will here only give a brief introduction to this powerful theory; for more
information see~\cite{rad:hc,kh:set,ju:av}. The idea in using wavefront sets is
to localize the distributions around the singular support and then to analyze
the directions in Fourierspace which causes these singularities. The advantage
of this approach is that only local concepts are used. In the following we will
give the definition of the wavefront set of a distribution and recall the main
result of Radzikowski's thesis.
\begin{definition}
  Let $(M,g)$ an $n$-dimensional manifold with associated cotangent bundle
  $T^{\ast}M$. Let $u\in\Dprime{M}$. A point $(x_0,k_0)\in T^{\ast}M\setminus
  \{0\}$ is called a regular directed point of $u\in\Dprime{M}$ if and only if for all
  $\lambda_0\in\Rn$ and for every function $\phi\in\mathcal{C}^{\infty}(M\times \Rn,
  \mathbf{R})$ with $d_x\phi(x_0,\lambda_0)=k_0$ there exists a neighborhood $U$ of $x_0$
  in $M$ and a neighborhood $\Lambda$ of $\lambda_0$ in $\Rn$, so that for all
  $\rho\in\mathcal{C}^{\infty}(U)$ and all $N\geq 0$ uniformly in $\lambda\in\Lambda$:
  \begin{equation}
    |\langle u,\rho \exp^{-i\tau\phi(\cdot,\lambda)}\rangle|=\text{o}(\tau^{-N})\quad
    \text{for}\quad \tau\rightarrow \infty                                                 
  \end{equation}
The wavefront set $\text{WF}(u)$ is then the  complement in $T^{\ast}M\setminus
\{0\}$ of the set of regular directed points of $u$.
\end{definition}
The wavefront set of a distribution $u$ has some remarkable properties:
\begin{enumerate}
  \item  The projection of WF$(u)$ on the first factor gives
             singsupp$(u)$
  \item  WF$(u)$ is a closed subset of $T^{\ast}M\setminus\{0\}$, since the
             complement of WF$(u)$ are open sets.
  \item  For all testfunctions $\phi$ with compact support it holds
             \begin{equation*}
               \text{WF}(\phi u)\subset \text{WF}(u)
             \end{equation*}
  \item  For a differential operator $D$ with $\mathcal{C}^{\infty}$ coefficients we have
             \begin{equation*}
               \text{WF}(Du)\subset \text{WF}(u)
             \end{equation*}
\end{enumerate}
The last relation means that the application of a linear differential operator
$D$ on $u$ does not enlarge the wavefront set of $u$.

For the wavefront set of a tensor product $u\otimes v$ of two distributions $u$
and $v$ on $M$ the following formula holds (see H\"ormander~\cite{hm:lpdo}):
\begin{equation}
  \begin{split}
  \text{WF}(u\otimes v)\subseteq & \ \text{WF}(u)\times\text{WF}(v)\\
                                 & \ \cup((\text{supp}(u)\times \{0\})\times \text{WF}(v))\\
                                 & \ \cup(\text{WF}(u)\times(\text{supp}(v)\times \{0\}))  
  \end{split}
\end{equation}

With the help of wavefront sets we are now able to characterize the Hadamard
states of the Dirac field. For this purpose we first look at the wavefront set
of the two-point function of the Klein-Gordon field. As was explained previously
we obtain the two-point function of the Dirac field through the application of
the adjoint Dirac-operator on the scalar two-point function. Due to the fact
that this procedure does not enlarge the wavefront set (since the Dirac-operator
is a linear differential operator) it is sufficient to determine the wavefront
set of the two-point function of the Klein-Gordon field. A completely local
description of Hadamard states of the Klein-Gordon field was given in
\cite{rad:hc}:
\begin{theorem}[Radzikowski, 1992]
  A quasifree state $\omega$ of the Klein-Gordon field on a globally hyperbolic spacetime
  $(M,g)$ is a Hadamard state, if and only if it's two-point function
  $\Lambda^{\omega}_{2,\text{KG}}$ has the following wavefront set:
  \begin{equation}
    \text{WF}(\Lambda^{\omega}_{2,\text{KG}})=\{(x_1,\xi_1;x_2,-\xi_2)\in T^{\ast}M^2\setminus
    \{0\};(x_1,\xi_1)\sim (x_2,\xi_2),\xi_1^0\geq 0\}                                      
  \end{equation}
\end{theorem}
\noindent Here $(x_1,\xi_1)\sim (x_2,\xi_2)$ means that $x_1$ and $x_2$ are connected by a
lightlike geodesic $\gamma$ in the way that $\xi_1^{\mu}$ is tangential to
$\gamma$ in $x_1$ and $\xi_2^{\mu}$ is $\xi_1^{\mu}$ parallely transported to
$x_2$. On the diagonal $(x_1,\xi_1)\sim (x_2,\xi_2)$ means that $\xi_1=\xi_2$
and $(\xi_1)^2=0$. We obtain the following result as a corollary to the above
theorem:
\begin{corollary}
  A state $\omega$ on $\mathcal{F}[\mathsf{H}]$ is a Hadamard state for the Dirac-field,
  if and only if it's two-point function $\Lambda_2^{\omega}$ has the following wavefront set
  \begin{equation}
    \text{WF}(\Lambda_2^{\omega})=\{(x_1,\xi_1;x_2,-\xi_2)\in T^{\ast}M^2\setminus
    \{0\};(x_1,\xi_1)\sim (x_2,\xi_2),\xi_1^0\geq 0\}                                      
  \end{equation}
\end{corollary}
\begin{proof}
This is an immediate consequence of property 4. above.
\renewcommand{\qed}{\hfill$\blacksquare$}
\end{proof}
One finds the
wavefront set of an arbitrary $m$-point function with the help of (50). When we
denote the kernel of $\Lambda_m^{\omega}$ by
$\Lambda_m^{\omega}(x_1,\ldots,x_m)$ we obtain
\begin{equation}
  \text{WF}(\Lambda^{\omega}_m(x_1,\ldots,x_m))\subseteq\bigcup_{\sigma}\bigoplus_{j=1}^n
  \text{WF}(\Lambda^{\sigma(j)}_2),\ n=m/2                                                 
\end{equation}
where $\Lambda^{\sigma(j)}_2$ is the two-point function in the variables
$x_{\sigma(j)},x_{\sigma(j+n)}$ viewed as a distribution over $M^m$, with
wavefront set
\begin{equation}
  \begin{split}
    \text{WF}(\Lambda^{\sigma(j)}_2) = & \ \{(x_1,0;\ldots;x_{\sigma(j)},k_{\sigma(j)};\ldots;
                                         x_{\sigma(j+n)},k_{\sigma(j+n)};\ldots;x_m,0)|       \\
                                       & \ (x_{\sigma(j)},k_{\sigma(j)};x_{\sigma(j+n)},
                                         k_{\sigma(j+n)})\in\text{WF}(\Lambda^{\sigma(j)}_2)\} \\
  \end{split}                                                                              
\end{equation}
That this expression is the wavefront set of $\Lambda_m^{\omega}$ is clear,
since $\Lambda_m^{\omega}$ is a sum over tensor products of two-point functions.

\section{Dirac-equation on Robertson-Walker spaces}
In this section we consider homogeneous and isotropic spacetimes. It is known
that the requirement of homogeneity and isotropy leads to the Robertson-Walker
spacetimes of the form $M=\mathbf{R}\times \Sigma^{\kappa}$, where
$\Sigma^{\kappa}$ is a homogeneous Riemannian manifold. The parameter $\kappa$
can take the values $+1,0$ or $-1$. Correspondingly $\Sigma^{\kappa}$ is a space
of constant positive, vanishing or constant negative curvature. We focus attention
to the Dirac-equation for which we construct explicit solutions by the method of
separation of variables due to Villalba and Percoco~\cite{vipe:sov}. We will use
these solutions in the next section to define adiabatic vacuum states of the
Dirac-field in this spacetime.

The metric in chronometrical coordinates $\tau,\chi,\theta,\varphi$ takes the
form
\begin{equation}
  ds^2=\text{e}^{\alpha(\tau)}[d\tau^2-d\chi^2-\xi^2(\chi)d\Omega^2]                       
\end{equation}
where $d\Omega^2$ is given by
\begin{equation}
  d\Omega^2=d\theta^2+\sin^2\theta d\varphi^2                                              
\end{equation}
$(\theta\in [0,\pi],\ \varphi\in [0,2\pi],\ \chi\in [0,\infty)\
\text{for}\ \kappa=0,-1\ \text{and}\ \chi\in [0,\pi)\ \text{for}\ \kappa=+1)$
and $\alpha(\tau)$ is a real-valued smooth function. The function $\xi(\chi)$ is
given by
\begin{equation}
  \xi(\chi) = \left\{ \begin{array}{ll}
                        \sin\chi  & \text{falls}\ \kappa=+1 \\
                        \chi      & \text{falls}\ \kappa=0 \\                              
                        \sinh\chi & \text{falls}\ \kappa=-1
                      \end{array}
              \right.
\end{equation}
To calculate the spinor connection we use a vierbein basis $e^a_{\mu}$ defined
by
\begin{equation}
  e^a_{\mu}e^b_{\nu}\eta_{ab}=g_{\mu\nu}                                                   
\end{equation}
The formula for the Christoffel symbols in this basis is given by
\begin{equation}
  \Ga{b}{ac}=e^b_{\nu}e^{\mu}_a \left( \partial_{\mu}e^{\nu}_c+
  e^{\lambda}_c\Ga{\nu}{\mu\lambda} \right)                                                
\end{equation}
In this expression $\Ga{\nu}{\mu\lambda}$ are the Christoffel-symbols in a
coordinate basis. Greek letters will always denote indices with respect to a
coordinate basis, whereas Latin indices correspond to a vierbein basis. We
choose the vierbein basis in the following way:
\begin{align}
  e^a_{\mu} & = \text{e}^{\alpha/2}(\delta^{a0}\delta_{\mu 0}+\delta^{a1}\delta_{\mu 1}+
                \xi(\chi)\delta^{a2}\delta_{\mu 2}+\xi(\chi)\sin\theta\delta^{a3}
                \delta_{\mu 3})\\                                                          
  e^{\nu}_b & = \text{e}^{-\alpha/2}(\delta^{\nu 0}\delta_{b0}+\delta^{\nu 1}\delta_{b1}+
                \xi^{-1}(\chi)\delta^{\nu 2}\delta_{b2}+\xi^{-1}(\chi)\sin^{-1}\theta
                \delta^{\nu 3}\delta_{b3})                                                 
\end{align}
With this input one finds after a straightforward computation the following
expression for the Christoffel symbols in vierbein basis:
\begin{equation}
  \begin{split}
  \Ga{b}{ac}= & \ \frac{1}{2}\text{e}^{-\alpha/2}\biggl\{ \dot{\alpha}\delta^{b0}\Bigl[
                \delta_{a1}\delta_{c1}+\delta_{a2}\delta_{c2}+\delta_{a3}\delta_{c3} \Bigr]\\
              & +\delta^{b1}\Bigl[ \delta_{a1}\delta_{c0}\dot{\alpha}-2\delta_{a2}\delta_{c2}
                \xi^{-1}(\chi)\xi^{\prime}(\chi)-2\delta_{a3}\delta_{c3}\xi^{-1}(\chi)
                \xi^{\prime}(\chi) \Bigr]\\
              & +\delta^{b2}\Bigl[ \delta_{a2}\delta_{c0}\dot{\alpha}+2\delta_{a2}\delta_{c1}
                \xi^{-1}(\chi)\xi^{\prime}(\chi)-2\delta_{a3}\delta_{c3}\xi^{-1}(\chi)\cot 
                \theta \Bigr]\\
              & +\delta^{b3}\Bigl[ \delta_{a3}\delta_{c0}\dot{\alpha}+2\delta_{a3}\delta_{c1}
                \xi^{-1}(\chi)\xi^{\prime}(\chi)+2\delta_{a3}\delta_{c2}\xi^{-1}(\chi)\cot
                \theta \Bigr] \bigg\}
  \end{split}
\end{equation}
The spinor connection is given by
\begin{equation}
  {{\sigma_a}^B}_A:=-\frac{1}{4}{\Gamma_{ca}}^{b}{{\gamma_{b}}^{B}}_{D}{\gamma^{cD}}_{A}   
\end{equation}
The $\gamma_a$ are the components of the spinor-tensor $\gamma$ w.r.t. the used
moving frame. The explicit expression for the spinor connection coefficients can
now be given:
\begin{equation}
  \begin{split}
  \sigma_a= & -\frac{1}{4}\text{e}^{-\alpha/2}\biggl\{ \dot{\alpha}\Bigl[ \delta_{a1}\gamma_0
              \gamma^1+\delta_{a2}\gamma_0\gamma^2+\delta_{a3}\gamma_0\gamma^3 \Bigr]\\
            & -2\delta_{a2}\gamma_1\gamma^2\xi^{-1}(\chi)\xi^{\prime}(\chi)-2\delta_{a3}\gamma_1
              \gamma^3\xi^{-1}(\chi)\xi^{\prime}(\chi)\\
            & -2\delta_{a3}\gamma_2\gamma^3\xi^{-1}(\chi)\cot\theta \biggr\}               
  \end{split}
\end{equation}
Here the anticommutation relations of the Dirac-matrices were used.

The Dirac-equation reads
\begin{equation}
  (-i\gamma^a\nabla_a+m)\Psi(x)=(-i\gamma^a e^{\mu}_a\nabla_{\mu}+m)\Psi(x)=0              
\end{equation}
Here the covariant derivative is lifted to the spinor bundle $DM$, as was
explained previously:
\begin{equation}
  \nabla_a=\partial_a+\sigma_a                                                             
\end{equation}
Inserting the expressions for the spinor connection coefficients and the
vierbeins we obtain the following explicit form of the Dirac-equation:
\begin{equation}
  \begin{split}
    \biggl\{ -i\text{e}^{-\alpha/2}\Bigl[ \gamma^0\Bigl(
    \partial_0-\frac{3}{4}\dot{\alpha}\Bigr)+
    \gamma^1\Bigl( \partial_1-\xi^{-1}(\chi)
    \xi^{\prime}(\chi) \Bigr)                                        + &     \\            
    \gamma^2\xi^{-1}(\chi)\Bigl( \partial_2-
    \frac{1}{2}\cot\theta \Bigr)+\gamma^3
    \xi^{-1}(\chi)\sin^{-1}\theta\partial_3 \Bigr]+m \biggr\}\Psi(x)   & = 0
  \end{split}
\end{equation}
In order to arrive at a more simpler equation we make the following transition
to the new spinor $\Phi$ defined by:
\begin{equation}
  \Phi(x)=\xi^{-1}(\chi)\sin^{-1/2}\theta\,\text{e}^{-3\alpha/4}\Psi(x),\ \chi\neq 0,\,
  \theta\neq 0,\pi,2\pi                                                                    
\end{equation}
The Dirac-equation for $\Phi$ now reads:
\begin{equation}
  \biggl\{ -i\text{e}^{-\alpha/2}\Bigl[ \gamma^0\partial_0+\gamma^1\partial_1+\gamma^2
  \xi^{-1}(\chi)\partial_2+\gamma^3\xi^{-1}(\chi)\sin^{-1}\theta\partial_3 \Bigr]+m \biggr\}
  \Phi(x)=0                                                                                
\end{equation}

We now want to solve this equation by the method of separation of variables.
This was already done in the paper by Villalba and Percoco~\cite{vipe:sov}, so
we shall outline here the main steps of this calculation only. All of the
missing details are completely straightforward.

First of all we decompose the Dirac-operator into two commuting differential
operators which can be dealt with separately. We use the Minkowski-space
Dirac-matrices in the following representation:
\begin{displaymath}
  \gamma^0=\begin{pmatrix} \mathbf{1}_2 & 0 \\ 0 & -\mathbf{1}_2\end{pmatrix}\quad
  \gamma^j=\begin{pmatrix} 0 & \sigma^j \\ -\sigma^j & 0\end{pmatrix},\quad j=1,2,3
\end{displaymath}
The Pauli-spinmatrices $\sigma^j$ are given by
\begin{displaymath}
  \sigma^1=\begin{pmatrix} 0 & 1 \\ 1 & 0\end{pmatrix}\quad
  \sigma^2=\begin{pmatrix} 0 & -i \\ i & 0\end{pmatrix}\quad
  \sigma^3=\begin{pmatrix} 1 & 0 \\ 0 & -1\end{pmatrix}
\end{displaymath}
The Dirac-equation can be written in the form:
\begin{equation}
  \left[ D^1(\theta,\varphi)+D^2_{\kappa}(\tau,\chi) \right]\tilde{\Phi}(x)=0              
\end{equation}
Here we have defined $\tilde{\Phi}=\gamma^1\gamma^0\Phi$ and the operators
$D^1(\theta,\varphi)$ and $D^2_{\kappa}(\tau,\chi)$ are given by:
\begin{align}
  D^1(\theta,\varphi)         & = \left[ \gamma^2\partial_2+\gamma^3\sin^{-1}\theta\partial_3
                                  \right]\gamma^1\gamma^0\\                                
  D^2_{\kappa}(\tau,\chi)     & = \xi(\chi)\left[ \gamma^0\partial_0+\gamma^1\partial_1+im
                                  \text{e}^{\alpha/2} \right]\gamma^1\gamma^0              
\end{align}
We observe that the operator $D^1(\theta,\varphi)$ is selfadjoint with respect
to the scalar product:
\begin{equation}
  (\Psi,\Phi):=\int_{0}^{\pi} d\theta\int_{0}^{2\pi} d\varphi\Psi_A(\theta,\varphi)^+
  \Phi^A(\theta,\varphi)                                                                   
\end{equation}
Now we have the following lemma:
\begin{lemma}
  $D^1$ and $D^2_{\kappa}$ are commuting operators on $\mathcal{C}^{\infty}_0(DM)$, i.e. we have
  \begin{equation}
    [\,D^1(\theta,\varphi)\,,\,D^2_{\kappa}(\tau,\chi)\,]=0                                        
  \end{equation}
\end{lemma}
\begin{proof}
We evaluate the product $D^1D^2_{\kappa}$:
\begin{align*}
  D^1D^2_{\kappa} = & \ \bigl[\gamma^2\gamma^1\gamma^0\partial_2+\gamma^3\gamma^1\gamma^0
                      \sin^{-1}\theta\partial_3 \bigr]\xi(\chi)\bigl[\gamma^0\partial_0+
                      \gamma^1\partial_1+im\text{e}^{\alpha/2} \bigr]\gamma^1\gamma^0\\
                  = & \ \gamma^1\gamma^0\bigl[\gamma^2\partial_2+\gamma^3\sin^{-1}\theta
                      \partial_3\bigr]\xi(\chi)\bigl[ \gamma^0\partial_0+\gamma^1\partial_1+
                      im\text{e}^{\alpha/2} \bigr]\gamma^1\gamma^0\\
                  = & -\gamma^1\gamma^0\xi(\chi)\bigl[ \gamma^0\partial_0+\gamma^1\partial_1-im
                      \text{e}^{\alpha/2} \bigr]\bigl[\gamma^2\partial_2+\gamma^3\sin^{-1}\theta
                      \partial_3 \bigr]\gamma^1\gamma^0\\
                  = & \ \xi(\chi)\bigl[ \gamma^0\partial_0+\gamma^1\partial_1
                      +im\text{e}^{\alpha/2}\bigr]\gamma^1\gamma^0\bigl[ \gamma^2\partial_2+
                      \gamma^3\sin^{-1}\theta\partial_3\bigr]\gamma^1\gamma^0\\
                  = & \ D^2_{\kappa}D^1
\end{align*}
\renewcommand{\qed}{\hfill$\blacksquare$}
\end{proof}
Because of this lemma it should be possible to find a common set of eigenspinors
for these two operators. Therefore we make the following ansatz to obtain a
solution to equation (70):
\begin{equation}
  D^1(\theta,\varphi)\tilde{\Phi}(x)=\lambda\tilde{\Phi}(x)=-D^2_{\kappa}(\tau,\chi)
  \tilde{\Phi}(x),\ \lambda\in\mathbf{C}                                                   
\end{equation}
First we look at the angle dependent part $D^1(\theta,\varphi)$:
\begin{equation}
  D^1(\theta,\varphi)\tilde{\Phi}=\lambda\tilde{\Phi}\Longleftrightarrow \Bigl\{ \gamma^2
  \partial_2+\gamma^3\sin^{-1}\theta\partial_3 \Bigr\}\gamma^1\gamma^0\tilde{\Phi}-\lambda
  \tilde{\Phi}=0                                                                           
\end{equation}
A solution to this equation can be obtained in terms of orthogonal polynomials,
i.e. we have the result
\begin{proposition}
  \begin{enumerate}
    \item  The eigenfunctions of the operator $D^1$ have the following form:
                \begin{equation}
                  \tilde{\Phi}(\theta,\varphi)=U^{-1}\Xi(\theta,\varphi)                   
                \end{equation}
                where $\Xi=\binom{\Xi_1}{\Xi_2}$ is a spinor with components
                \begin{equation}
                  \Xi_1(\theta,\varphi)=\binom{C^1_{nl}(\theta,\varphi)}
                  {C^2_{nl}(\theta,\varphi)},\quad
                  \Xi_2(\theta,\varphi)=\binom{C^1_{nl}(\theta,\varphi)}
                  {-C^2_{nl}(\theta,\varphi)}                                              
                \end{equation}
                with $l\in \mathbf{N},\ n=\frac{1}{2}(2m+1),\,m\in\mathbf{N}$
                and scalar valued functions
                \begin{align}
                  C^1_{nl}(\theta,\varphi) & = \sin^n\theta\cos(\theta/2)G_{nl}(\cos\theta)
                                               W_n(\varphi)\\                              
                  C^2_{nl}(\theta,\varphi) & = \sin^n\theta\sin(\theta/2)F_{nl}(\cos\theta)
                                               W_n(\varphi)                                
                \end{align}
                $G_{nl}$ and $F_{nl}$ are the Jacobi-polynomials $P_l^{(a,b)}$ in the variable
                $\cos\theta$ of order $l$, i.e.
                \begin{align}
                  G_{nl}(\cos\theta) & = \frac{1}{\sqrt{N_{nl}}}P_l^{(\beta_1,\beta_2)}
                                         (\cos\theta) \\                                   
                  F_{nl}(\cos\theta) & = (-1)^l\frac{1}{\sqrt{N_{nl}}}P_l^{(\beta_2,\beta_1)}
                                         (\cos\theta)                                      
                \end{align}
                The parameters $\beta_1$ and $\beta_2$ are given by
                \begin{align*}
                  \beta_1 & = n-\frac{1}{2} \\
                  \beta_2 & = n+\frac{1}{2}
                \end{align*}
                and $N_{nl}$ is a normalization constant. The scalar valued function
                $W_n$ is given by
                \begin{equation}
                  W_n(\varphi)=\exp(in\varphi)                                             
                \end{equation}
                And the unitary matrix $U$ is given by
                \begin{equation}
                  U\equiv\frac{1}{2}\left( \gamma^2\gamma^1+\gamma^3\gamma^2+\gamma^1\gamma^3+
                  \mathbf{1}_4 \right)                                                     
                \end{equation}
    \item  The eigenvalues $\lambda$ are given by
                \begin{equation}
                  \lambda=l+n+\frac{1}{2}\in\mathbf{N}\setminus\{0\}                       
                \end{equation}
  \end{enumerate}
  The Jacobi-polynomials are defined in the proof.
\end{proposition}

\begin{proof}
Eq.(76) is a system of partial differential equations in two variables and the aim of the
following calculation is to separate these two variables. The matrix $U$
transforms the Dirac-matrices in the following way:
\begin{displaymath}
  U\gamma^0U^{-1}=\gamma^0,\ U\gamma^1U^{-1}=\gamma^3,\ U\gamma^2U^{-1}=\gamma^1,\
  U\gamma^3U^{-1}=\gamma^2
\end{displaymath}
The application of $U$ to equation (76) then gives
\begin{equation}
  \Bigl\{ \gamma^1\partial_2+\gamma^2\sin^{-1}\theta\partial_3 \Bigr\}                     
  \gamma^3\gamma^0\Xi-\lambda\Xi=0
\end{equation}
Inserting the above form of the Dirac-matrices we obtain
\begin{equation}
  \biggl\{ \begin{pmatrix} \sigma^2 & 0 \\ 0 & -\sigma^2\end{pmatrix}\partial_2+
           \begin{pmatrix} -\sigma^1 & 0 \\ 0 & \sigma^1\end{pmatrix}\sin^{-1}\theta\partial_3
  \biggr\}\binom{\Xi_1}{\Xi_2}-\lambda\binom{\Xi_1}{\Xi_2}=0                               
\end{equation}
The variable $\varphi$ enters this system of
equations in the derivative operators only. Therefore we make the following
ansatz for the spinor $\Xi$:
\begin{displaymath}
  \Xi_1=a_1(\tau)b_1(\chi)\binom{\chi_1(\theta)}{\chi_2(\theta)}W(\varphi),\quad
  \Xi_2=a_3(\tau)b_3(\chi)\binom{\chi_3(\theta)}{\chi_4(\theta)}W(\varphi)
\end{displaymath}
This is the most general ansatz for the spinor $\Xi$, for which one can separate
the variables $\theta$ and $\varphi$ in eq.(87). This form is also compatible
with the structure of the operator $D^2_{\kappa}(\tau,\chi)$ as we will see
later. When we insert this into equation (87), we see that we have
\begin{displaymath}
  \chi_1=\chi_3\quad\chi_2=-\chi_4
\end{displaymath}
so we obtain only two coupled linear independent partial differential equations in the
variables $\theta$ and $\varphi$. Now we can separate this system in the
following way:
\begin{align}
  \sin\theta\chi_2^{-1}(i\lambda\chi_1-
  \partial_{\theta}\chi_2)                & = n=-iW^{-1}\partial_{\varphi}W \\             
  \sin\theta\chi_1^{-1}(i\lambda\chi_2+
  \partial_{\theta}\chi_1)                & = n=-iW^{-1}\partial_{\varphi}W                
\end{align}

The solution for $W$ reads:
\begin{equation}
   W_n(\varphi)=\exp(in\varphi)                                                            
\end{equation}
We demand the spinor $\Psi$ to change the sign when rotated about an angle
$2\pi$ in the direction $\varphi$, so $n$ varies in the range
$n=\pm\frac{1}{2},\pm\frac{3}{2},\pm\frac{5}{2},\ldots$.

Next we have to solve the remaining system of equations in the variable
$\theta$:
\begin{align}
  \Bigl( \partial_{\theta}+\frac{n}{\sin\theta} \Bigr)\chi_2-\tilde{\lambda}\chi_1 & = 0\\ 
  \Bigl( \partial_{\theta}-\frac{n}{\sin\theta} \Bigr)\chi_1+\tilde{\lambda}\chi_2 & = 0   
\end{align}
with $\tilde{\lambda}:=i\lambda$. We make for $\chi_1$ and $\chi_2$ the ansatz
\begin{align}
  \chi_{1n}(\theta) & = \sin^n\theta\cos(\theta/2)g(\theta)\\                              
  \chi_{2n}(\theta) & = \sin^n\theta\sin(\theta/2)f(\theta)                                
\end{align}
where $f$ and $g$ are unknown scalar functions, still to be determined.
Inserting this into equation (92) we obtain after a little manipulation
\begin{equation}
  \begin{split}
    \cos(\theta/2)\Bigl[ ng(\theta)\Bigl( \cos\theta-1 \Bigr)-\sin^2\theta
    \dot{g}(\theta) \Bigr]                                                  + &     \\
    \sin(\theta/2)\Bigl[ \sin\theta\Bigl( \tilde{\lambda} f(\theta)-
    \frac{1}{2}g(\theta) \Bigr) \Bigr]                                        & = 0        
  \end{split}
\end{equation}
where
\begin{equation}
  g^{\prime}(\theta)=\frac{dg}{d\cos\theta}\frac{d\cos\theta}{d\theta}\equiv-\dot{g}(\theta)
  \sin\theta                                                                               
\end{equation}
With the identity
\begin{equation}
  \cot(\theta/2)=\frac{1+\cos\theta}{\sin\theta}                                           
\end{equation}
we finally obtain
\begin{equation}
  -ng(\theta)-(1+\cos\theta)\dot{g}(\theta)+\tilde{\lambda} f(\theta)
  -\frac{1}{2}g(\theta)=0                                                                  
\end{equation}
With equation (91) we proceed in the same way and find
\begin{equation}
  nf(\theta)-(1-\cos\theta)\dot{f}(\theta)+\frac{1}{2}f(\theta)
  -\tilde{\lambda}g(\theta)=0                                                              
\end{equation}
With $x:=\cos\theta$ these two equations become
\begin{align}
  (1+x)\frac{dG}{dx}+\left( \frac{1}{2}+n \right)G & = \tilde{\lambda}F \\                
  (x-1)\frac{dF}{dx}+\left( \frac{1}{2}+n \right)F & = \tilde{\lambda}G                   
\end{align}
The functions $F$ and $G$ are defined by
\begin{align}
  F(x) & := f(\theta) \\                                                                  
  G(x) & := g(\theta)                                                                     
\end{align}
When we now insert eq.(100) into eq.(101) we obtain a differential equation
which is solved by the Jacobi-polynomials~\cite{abst:hmf}:
\begin{equation}
  (1-x^2)\ddot{G}+(\beta_2-\beta_1-(\beta_1+\beta_2+2)x)\dot{G}
  +l(l+\beta_1+\beta_2+1)G=0                                                              
\end{equation}
Here we have made the definitions:
\begin{align*}
  \beta_1 & := n-\frac{1}{2} \\
  \beta_2 & := n+\frac{1}{2} \\
  l       & := |\tilde{\lambda}|-|n|-\frac{1}{2}
\end{align*}
The Jacobi-polynomials of order $l$ are defined by
\begin{align*}
  P^{(\beta_1,\beta_2)}_l(x) & = 2^{-l}\sum_{m=0}^lc_m(\beta_1,\beta_2)g_m(x),\,l\in\mathbf{N}
                                 \ \text{where} \\
  c_m(\beta_1,\beta_2)       & = \binom{l+\beta_1}{m}\binom{l+\beta_2}{l-m} \\
  g_m(x)                     & = (x-1)^{l-m}(x+1)^m
\end{align*}
The parameters $\beta_1,\beta_2$ are restricted to the values
\begin{equation*}
  \beta_1>-1,\quad\beta_2>-1
\end{equation*}
This restrict the values of the parameter $n$ further to
$n=\frac{1}{2}(2m+1),\,m\in\mathbf{N}$ The solution for $G$ now reads
\begin{equation}
  G_{nl}(x)=\frac{1}{\sqrt{N_{nl}}}P^{(\beta_1,\beta_2)}_l(x)                             
\end{equation}
where $N_{nl}$ is a normalization constant. The eigenvalues $|\tilde{\lambda}|$,
which are related to the order of the Jacobi-polynomials $l$ and to $n$ via
$l=|\tilde{\lambda}|-n-\frac{1}{2}$ must be integer, since $l$ is an integer.
With $G$ we easily obtain the solution for $F$:
\begin{equation}
  F_{nl}(x)=\frac{1}{\sqrt{N_{nl}}}(-1)^lP^{(\beta_2,\beta_1)}_l(x)                       
\end{equation}
\renewcommand{\qed}{\hfill$\blacksquare$}
\end{proof}

If we choose $N_{nl}$ to be
\begin{equation}
  N_{nl}=\frac{2^{2n+3}\pi\Gamma(l+n+1/2)\Gamma(l+n+3/2)}{(2l+2n+1)
         \Gamma(l+2n+1)l!}                                                                
\end{equation}
then the eigenfunctions $\tilde{\Phi}(\theta,\varphi)$ of $D^1$ are orthonormal,
i.e. we have the relation:
\begin{equation}
  \int_{0}^{\pi}d\theta\int_{0}^{2\pi}d\varphi \tilde{\Phi}^+_{nl}(\theta,\varphi)_A      
  \tilde{\Phi}_{n^{\prime}l^{\prime}}(\theta,\varphi)^A=\delta_{nn^{\prime}}\delta_{ll^{\prime}}
\end{equation}

We now consider the operator $D^2_{\kappa}(\tau,\chi)$. Equation (75) reads
\begin{align*}
  D^2_{\kappa}(\tau,\chi)\tilde{\Phi}+\lambda\tilde{\Phi}     & =   \\
  \Bigl[ \gamma^0\partial_0+\gamma^1\partial_1+im
  \text{e}^{\alpha/2} \Bigr]\gamma^1\gamma^0\tilde{\Phi}+
  \tilde{\lambda}\xi^{-1}\tilde{\Phi}                         & = 0
\end{align*}
After the application of $U$,  this equation becomes
\begin{equation}
  \Bigl[ -\gamma^3\partial_0-\gamma^0\partial_1+\gamma^3\gamma^0im\text{e}^{\alpha/2}+
  \tilde{\lambda}\xi^{-1} \Bigr]\Xi=0                                                     
\end{equation}
Inserting the Dirac-matrices we obtain the following system of equations:
\begin{equation}
  \begin{split}
    -\sigma^3\partial_0\Xi_2-\partial_1\Xi_1-\sigma^3im\text{e}^{\alpha/2}\Xi_2+
    \tilde{\lambda}\xi^{-1}\Xi_1                                                      & = 0 \\
    \sigma^3\partial_0\Xi_1+\partial_1\Xi_2-\sigma^3im\text{e}^{\alpha/2}\Xi_1+           
    \tilde{\lambda}\xi^{-1}\Xi_2                                                      & = 0
  \end{split}
\end{equation}
As in the angle dependent case we are looking for a separation of the variables
$\tau$ and $\varphi$. Therefore we enlarge our ansatz for the spinor $\Xi$ in
the following way:
\begin{align}
  \Xi_1 & = \binom{a_1(\tau)b_1(\chi)C^1_{nl}(\theta,\varphi)}{a_2(\tau)b_2(\chi)
            C^2_{nl}(\theta,\varphi)}\\                                                   
  \Xi_2 & = \binom{a_3(\tau)b_3(\chi)C^1_{nl}(\theta,\varphi)}{-a_4(\tau)b_4(\chi)
            C^2_{nl}(\theta,\varphi)}                                                     
\end{align}
where $a_i,b_j,\ i,j=1,\ldots,4$ are unknown scalar valued functions, to be
determined in the following. As was mentioned earlier in the proof of
proposition 1, to be compatible with both operators $D^1$ and $D^2_{\kappa}$ we
must have
\begin{alignat*}{2}
  a_1 & = a_2, & \qquad b_1 & = b_2 \\
  a_3 & = a_4, & \qquad b_3 & = b_4
\end{alignat*}
This is indeed the case, as can be seen when eqs.(111), (112) are inserted in
(110). Therefore we have to solve only the reduced system
\begin{align}
  a_3^{-1}\partial_0a_1+b_1^{-1}\partial_1b_3
  -im\text{e}^{\alpha/2}a_1a_3^{-1}+\tilde{\lambda}\xi^{-1} b_3b_1^{-1} & = 0\\           
  -a_1^{-1}\partial_0a_3-b_3^{-1}\partial_1b_1
  -im\text{e}^{\alpha/2}a_3a_1^{-1}+\tilde{\lambda}\xi^{-1} b_1b_3^{-1} & = 0             
\end{align}
Introducing the two constants $z_1,\,z_2\in\mathbf{C}$ we obtain for the
variable $\tau$:
\begin{align}
  \Bigl( \frac{d}{d\tau}-im\text{e}^{\alpha/2} \Bigr)a_1(\tau) & = z_1a_3(\tau)\\         
  \Bigl( \frac{d}{d\tau}+im\text{e}^{\alpha/2} \Bigr)a_3(\tau) & = z_2a_1(\tau)           
\end{align}
And for the variable $\chi$:
\begin{align}
  \Bigl( \frac{d}{d\chi}+\tilde{\lambda}\xi^{-1}(\chi) \Bigr)b_3(\chi) & = -z_1b_1(\chi)\\ 
  \Bigl( \frac{d}{d\chi}-\tilde{\lambda}\xi^{-1}(\chi) \Bigr)b_1(\chi) & = -z_2b_3(\chi)   
\end{align}

The equations in the radial coordinate $\chi$ can be solved by hypergeometric
functions.
\begin{proposition}
  For $\kappa=0$ the solutions to the eqn.(117) and (118) can be expressed through the functions
  \begin{align}
    b_1(\chi) & = B_{1w}(y)=B_k(d_1y\,\mathsf{j}_k(y)+d_2y\,\mathsf{y}_k(y))      \\      
    b_3(\chi) & = B_{3w}(y)=B_k(d_1y\,\mathsf{j}_{k-1}(y)+d_2y\,\mathsf{y}_{k-1}(y))      
  \end{align}
  where $d_1,d_2\in\mathbf{R},\ k\in\mathbf{Z},\ y=\pm\sqrt{w}\chi,\ \text{with}\
  w\in \mathbf{R}_+$
\end{proposition}

\begin{proof}
First of all we take $\kappa=0$ in (117) and (118) and combine them to the
single equation of second order
\begin{equation}
  b_1^{\prime\prime}-b_1\left( \tilde{\lambda}^2\chi^{-2}-\tilde{\lambda}\chi^{-2}+z
  \right)=0                                                                               
\end{equation}
where  $z=z_1z_2$ and a prime denotes a derivative with respect to $\chi$. For
$z\in\mathbf{R}_{-}:=\{z\in\mathbf{R}|z<0\}$\footnote{On physical grounds there
are only negative values for $z$ of interest, since it follows from the
treatment of the equations (115) and (116) that $-z$ is the absolute value
squared of the wavevector $\vec{k}$.} we set $-z=:w>0$ and define a new variable
$y$ by $y:=\pm\sqrt{w}\chi$. The derivative w.r.t $y$ is denoted by a dot. Then
equation (121) becomes
\begin{equation}
  y^2\ddot{B}_1(y)+B_1(y)[\tilde{\lambda}(1-\tilde{\lambda})+y^2]=0                       
\end{equation}
where the function $B_1$ is defined by $B_1(y):=b_1(\chi)$. If we finally set
$\tilde{\lambda}=-k=-k(l,n)$ we obtain the so called Riccati-Bessel differential
equation~\cite{abst:hmf}:
\begin{equation}
  y^2\ddot{B}_1(y)+B_1(y)[y^2-k(k+1)]=0                                                   
\end{equation}
The solutions to this equation are the spherical Bessel-functions of the first
and second kind $\mathsf{j}_k(y)$ and $\mathsf{y}_k(y)$ respectively. The
general solution is a linear combination of these two:
\begin{equation}
  B_{1w}(y)=B_k\left( d_1y\,\mathsf{j}_k(y)+d_2y\,\mathsf{y}_k(y) \right),
  \quad k\in\mathbf{Z},\ d_1,d_2\in\mathbf{R}                                             
\end{equation}
where $B_k$ is a normalization constant. The solution for $b_{3w}$ is, expressed
through the function $B_{3w}(y):=b_{3w}(\chi)$,
\begin{equation}
  B_{3w}(y)=B_k\sqrt{\frac{\pi}{2}}y^{-1/2}\left( d_1\mathsf{J}_{k-1/2}(y)+
  d_2\mathsf{Y}_{k-1/2}(y) \right)                                                        
\end{equation}
where the $\mathsf{J}_{k+1/2}(y)$ and $\mathsf{Y}_{k+1/2}(y)$ are the usual
Bessel-functions of rational order of first resp. second kind. They are related
to the spherical ones by
\begin{equation}
  \mathsf{j}_k(y)=\sqrt{\frac{\pi}{2}}y^{-1/2}\mathsf{J}_{k+1/2}(y),\quad
  \mathsf{y}_k(y)=\sqrt{\frac{\pi}{2}}y^{-1/2}\mathsf{Y}_{k+1/2}(y)
\end{equation}
\renewcommand{\qed}{\hfill$\blacksquare$}
\end{proof}

We summarize the above calculation in the following two statements
\begin{proposition}
  The solution to the Dirac-equation eq.(67) has the following spinor structure:
  \begin{equation}
    \Psi_{wnl}(x)=-\text{e}^{3\alpha/4}\chi\sin^{1/2}\theta M\Xi_{wnl}(x)                 
  \end{equation}
  where $M$ is the matrix $M=\gamma^0\gamma^1U^{-1}$ and $\Xi_{wnl}(x)$ is given by the
  equations (111) and (112).
\end{proposition}
\begin{proposition}
   The spatial part of the solution of the Dirac-equation (67) in the coordinates $\chi,
   \theta,\varphi$ on a flat Robertson-Walker spacetime has the form:
   \begin{equation}
     \mathbf{\Psi}(\vec{x})=\int dw\sum_{l,n}\left( a_{nl}(w)\Psi_{wnl}(\vec{x})
     +a_{nl}^{\ast}(w)\overline{\Psi_{wnl}(\vec{x})} \right)                             
   \end{equation}
   with basis $\Psi_{wnl}$ given by
   \begin{equation*}
     \Psi_{wnl}(\vec{x})=-\chi\sin^{1/2}\theta M\Xi_{wnl}(\vec{x})
   \end{equation*}
   where
   \begin{equation*}
     \Xi_{wnl}(\vec{x})=\binom{\Xi_{1wnl}(\vec{x})}{\Xi_{2wnl}(\vec{x})}
   \end{equation*}
   with
   \begin{align*}
     \Xi_{1wnl}(\vec{x}) & = b_{1w}(\chi)\binom{C^1_{nl}(\theta,\varphi)}{C^2_{nl}(\theta,\varphi)}\\
     \Xi_{2wnl}(\vec{x}) & = b_{3w}(\chi)\binom{C^1_{nl}(\theta,\varphi)}{-C^2_{nl}(\theta,\varphi)}
   \end{align*}
\end{proposition}

The coefficient function $a_{nl}(w)$ in eq.(128) must be chosen with compact
support to make the integral convergent. The solution to the equations (117) and
(118) in the cases $\kappa=\pm 1$ can be given equally in terms of more advanced
hypergeometric functions but there is only little to learn of these formulas, so
we skip them here.

The treatment of the equations (115) and (116) requires the specification of the
function $\alpha$, which corresponds to choosing a certain universe model.

\section{Adiabatic vacuum states}
We come now to the discussion of adiabatic vacuum states on
$\mathcal{F}[\mathsf{H}]$. Associated with each homogeneous space
$\Sigma^{\kappa}$ there is a group of isometries $G^{\kappa}$. This group acts
through $g(t,\vec{x})=(t,g\vec{x}),\ g\in G^{\kappa}$ on $M$. In the case of
constant positive curvature $(\kappa=+1)$ this is the rotation group
$\text{SO}(4)$. If we take flat Cauchy-surfaces we have
$G^{\kappa}=\text{E}(3)$. In the case of constant negative curvature
$(\kappa=-1)$ the symmetries are the proper orthochronous
Lorentz-transformations $\mathcal{L}^{\uparrow}_+$. See ref.~\cite{ldro:lqav}
for more details.

We now want to characterize certain states on the CAR-Algebra
$\mathcal{F}[\mathsf{H}]$ which are invariant under the action of the
transformation group $G^{\kappa}$.  The unitary representation $U$ of
$G^{\kappa}$ on $\mathcal{C}^{\infty}_0(DM)$, defined by
\begin{equation}
  (U(g)f)(x):=f(g^{-1}x),\ g\in G^{\kappa}                                                
\end{equation}
commutes with the Dirac-propagator for spinors
$S:\mathcal{C}^{\infty}_0(DM)\rightarrow \mathcal{C}^{\infty}(DM)$. In addition
to that it holds that the operators $U(g)$ commute with the antiunitary
involution $\Gamma$ on $\mathsf{H}$. In this way they form a group of
Bogoliubov-transformations on $\mathsf{H}$. Induced by these transformations
there are automorphisms $\alpha_g$ of $\mathcal{F}[\mathsf{H}]$, i.e.
\begin{equation}
  \alpha_g(B(f))=B(U(g)f),\ g\in G^{\kappa}                                               
\end{equation}
These so called Bogoliubov-automorphisms are now used to single out homogeneous
and isotropic states of the Dirac-field. A state $\omega$ on
$\mathcal{F}[\mathsf{H}]$ is said to be homogeneous and isotropic, if and only
if it is invariant under all Bogoliubov-automorphisms $\alpha_g$, i.e.
\begin{equation}
  \omega\circ\alpha_g=\omega,\ \forall g\in G^{\kappa}                                    
\end{equation}
The restriction to quasifree states makes it possible to formulate an equivalent
condition on the two-point function
\begin{equation}
  \Lambda^{\omega}_2(f_1,f_2):=\omega(B(f_1)^{\ast}B(f_2))                                
\end{equation}
\begin{definition}
  A quasifree state $\omega$ on $\mathcal{F}[\mathsf{H}]$ is called homogeneous and
  isotropic, if and only if it's two-point function satisfies
  \begin{equation}
    \Lambda^{\omega}_2(U(g)f_1,U(g)f_2)=\Lambda^{\omega}_2(f_1,f_2),\ f_1,f_2\in
    \mathsf{H},\ g\in G^{\kappa}                                                          
  \end{equation}
\end{definition}

We now introduce adiabatic vacuum states for the Dirac-field on a
Robertson-Walker spacetime. For the Klein-Gordon-field these states were invented
by L\"uders and Roberts~\cite{ldro:lqav}, motivated by Parkers requirement that
the particle production by the expanding universe should be
minimal~\cite{pa:pceu1}. It has been shown that adiabatic vacuum states for the
Klein-Gordon-field are the physically correct states, in the sense that they
have the singularity structure of Hadamard-form~\cite{ju:av}. This fact
motivates the same procedure in the case of the Dirac-field. First of all we
write the field operator according to proposition 4 as:
\begin{equation}
  \hat{\Psi}(x)=\int dw\sum_{l,n}\left[ \hat{a}_{nl}(w)\Psi_{wnl}(x)+
  \hat{a}_{nl}^+(w)\overline{\Psi_{wnl}(x)} \right]                                       
\end{equation}
where $\hat{a}_{nl}(w),\hat{a}_{nl}(w)^+$ are annihilation and creation
operators on the antisymmetric Fock-space over $\mathsf{H}$. They have
anticommutation relations
\begin{equation}
  \{\,\hat{a}_{nl}(w)\,,\,\hat{a}_{n^{\prime}l^{\prime}}(w^{\prime})^{\ast}\,\}=
  \delta(w-w^{\prime})\delta_{nn^{\prime}}\delta_{ll^{\prime}}
\end{equation}                                                                            
We now consider the equations (115) and (116) for the time dependent part.
Combining them we get
\begin{equation}
  \Bigl( \frac{d^2}{d\tau^2}-i\frac{\dot{\alpha}}{2}m\text{e}^{\alpha/2}+
  m^2\text{e}^{\alpha} \Bigr)a_{1w}(\tau)=-wa_{1w}(\tau)                                  
\end{equation}
In analogy to the Klein-Gordon case we make a WKB-type ansatz for a solution to
equation (136)
\begin{equation}
  a_{1w}(\tau)=\text{e}^{-3\alpha/4}\left( 2\Omega_w(\tau) \right)^{-1/2}
  \exp\left[ i\int_0^{\tau}d\tau^{\prime}\Omega_w(\tau^{\prime}) \right]                  
\end{equation}
where $\Omega_w$ is a complex valued smooth function with positive imaginary
part. If we insert this in (136) we get an equation for $\Omega_w$:
\begin{equation}
  \begin{split}
    \Omega_w(\tau)^2 = & \ \frac{9}{16}\dot{\alpha}^2+\frac{3}{4}\dot{\alpha}
                         \frac{\dot{\Omega}_w(\tau)}{\Omega_w(\tau)}-\frac{3i}{2}
                         \dot{\alpha}\Omega_w(\tau)+\frac{1}{4}\Bigl( \frac{
                         \dot{\Omega}_w(\tau)}{\Omega_w(\tau)} \Bigr)^2-\frac{3}{4}       
                         \ddot{\alpha} \\
                       & - \frac{1}{2}\frac{\ddot{\Omega}_w(\tau)}{\Omega_w(\tau)}+
                         \frac{1}{2}\frac{\dot{\Omega}_w(\tau)}{\Omega_w(\tau)^2}-
                         \frac{i}{2}\dot{\alpha}m\text{e}^{\alpha/2}+m^2\text{e}^{\alpha}+w
  \end{split}
\end{equation}
In a slowly varying universe we expect the derivative terms to be small, so we
try an iterative solution to this equation:
\begin{align*}
  \bigl( \Omega^{(0)}_w(\tau) \bigr)^2   = & \ w+m^2\text{e}^{\alpha} \\
  \bigl( \Omega^{(r+1)}_w(\tau) \bigr)^2 = & \ w+m^2\text{e}^{\alpha}-\frac{i}{2}\dot{\alpha}
                                               m\text{e}^{\alpha/2}+\frac{9}{16}\dot{\alpha}^2-
                                               \frac{3}{4}\ddot{\alpha}+\frac{3}{4}\dot{\alpha}
                                               \frac{\dot{\Omega}^{(r)}_w(\tau)}{\Omega^{(r)}_w
                                               (\tau)} \\
                                           & - \frac{3i}{2}\dot{\alpha}\Omega^{(r)}_w(\tau)
                                               +\frac{1}{4}\Bigl( \frac{\dot{\Omega}^{(r)}_w
                                               (\tau)}{\Omega^{(r)}_w(\tau)} \Bigr)^2-
                                               \frac{1}{2}\frac{\ddot{\Omega}^{(r)}_w(\tau)}
                                               {\Omega^{(r)}_w(\tau)}+\frac{1}{2}\frac{\dot{
                                               \Omega}^{(r)}_w(\tau)}{\bigl( \Omega^{(r)}_w
                                               (\tau)\bigr)^2}
\end{align*}
The only freedom we have in the choice of a homogeneous and isotropic quasifree
state is to specify initial data for the functions $a_{1w}$ and $a_{3w}$.
Therefore we make the following definition:
\begin{definition}
  An adiabatic vacuum state of the Dirac-field of order $r$ on a flat Robertson-Walker
  spacetime is specified by the following initial data for the function $a_{1w}$:
  \begin{align}
    a_{1w}(\tau)       = & \ W^{(r)}_w(\tau) \\                                           
    \dot{a}_{1w}(\tau) = & \ \dot{W}^{(r)}_w(\tau),\ \text{where}                         
  \end{align}
  \begin{equation}
    W^{(r)}_w(\tau):=\text{e}^{-3\alpha/4}\bigl( 2\Omega^{(r)}_w(\tau)
    \bigr)^{-1/2}\exp\left[ i\int_0^{\tau}d\tau^{\prime}\Omega^{(r)}_w                    
    (\tau^{\prime}) \right]
  \end{equation}
\end{definition}
The function $a_{3w}$ is then fixed by eq.(115). Now one calculates the
two-point distribution to be
\begin{equation}
  \begin{split}
    \Lambda_2(x_1,x_2) = & \ \langle\Omega,\hat{\Psi}(x_1)\hat{\Psi}(x_2)\Omega\rangle \\
                         & \ \int dw\sum_{nl}\Psi_{wnl}(x_1)\overline{\Psi_{wnl}(x_2)}    
  \end{split}
\end{equation}
where the anticommutation relations eq.(135) were used. To decide weather the
quasifree state associated with this two-point distribution is an Hadamard-state
we must look at the short distance behaviour of the expression (142). If we use
the Taylor expansions of the Bessel-functions we arrive at a form which agrees
with the one in definition 1. So we conjecture that we obtain in this way a
Hadamard-state.

\section{Appendix}
In this appendix we define bi-spinors in curved spacetime and introduce the
relevant notions of the causal structure of this spacetime for the definition of
Hadamard states for the Dirac-field. In the following our spacetime is a smooth
$\mathcal{C}^{\infty}$-manifold on which there is defined a Lorentz metric $g$.
In addition to that we assume that $(M,g)$ is globally hyperbolic.

Let $N$ be a open subset of $M$. Further let $DN\boxtimes D^{\ast}N$ be the
outer tensor product of the  bundles $DN$ and $D^{\ast}N$. This is the bundle
over the product manifold $N\times N$. the fibres at $(p,p')\in N\times N$ are
$(DN)_p\otimes (D^{\ast}N)_{p'}$. $\pi$ is the product-projection, i.e.
$\pi((p,f),(p',f'))=(p,p')$. Smooth sections in $DN\boxtimes D^{\ast}N$ are
called bi-spinors. With the choice of vierbeins $(E_A)$ at $p$ and $({E'}^{B'})$
at $p'$, we construct a new vierbein  $(q,q')\mapsto {(E\boxtimes
E')_A}^{B'}(q,q'):=E_A(q)\otimes {E'}^{B'}(q'),\,A,B'=0,\ldots,3$ in
$DN\boxtimes D^{\ast}N$ at $(p,p')$. Smooth sections $v\in
\mathcal{C}^{\infty}(DN\boxtimes D^{\ast}N)$ are given by
\begin{equation}
  v(q,q')={v^A}_{B'}(q,q'){(E\boxtimes E')_A}^{B'}(q,q')
\end{equation}
with $\mathcal{C}^{\infty}$-functions ${v^A}_{B'}$.

For the singularity structure of the two-point function to be well defined we
need the notion of a causal normal neighborhood of a Cauchy-surface $\Sigma$,
which we explain now. An open neighborhood $N$ of a Cauchy-surface $\Sigma$ in
$(M,g)$ is called a causal normal neighborhood for $\Sigma$, if for every choice
of $p,q\in N$ with $p\in J^+(q)$ there is a convex normal
neighborhood~\cite{wa:gr} $\mathcal{R}\subset M$, so that $J^-(p)\cap
J^+(q)\subset \mathcal{R}$. It was shown by Kay and Wald~\cite{kawa:tutp} that
to every Cauchy-surface there exists a causal normal neighborhood. Now let $X$
be the set of points $(p,q)\in M\times M$, such that $p$ and $q$ can be joined
by a causal curve with $J^+(p)\cap J^-(q)$, if $q\in J^+(p)$ or $J^-(p)\cap
J^+(q)$, if $p\in J^+(q)$ contained in a convex normal neighborhood. The square
of the geodesic distance $s(p,q)$ of $p$ and $q$ is then well defined and smooth
in $X\subset M\times M$.

Let $N$ be a causal normal neighborhood of a Cauchy-surface $\Sigma$. A smooth
function $\chi$ on $N\times N$ is called a $N$-regularizing function, if it has
the following properties: There is an open neighborhood $Y\subset N\times N$ of
the set of causally connected points in $N$, such that $\overline{Y}\subset X$
and $\chi \equiv 1$ on $Y$ and $\chi\equiv 0$ outside of $X$. Then $\chi$ is
well defined, since $Y$ and the complement of $X$ in $N\times N$ are disjoined
closed subsets of $N\times N$. It can be shown that such a $N$-regularizing
function always exists~\cite{rad:hc}.

\providecommand{\bysame}{\leavevmode\hbox to3em{\hrulefill}\thinspace}

\end{document}